\documentclass[preprintnumbers,article,amsmath,amssymb,floatfix,10pt,prd,onecolumn,superscriptaddress,nofootinbib]{revtex4-2}

\usepackage{titlesec}
\usepackage[toc]{appendix}
\usepackage{caption}
\titleformat*{\section}{\LARGE\bfseries}
\titleformat*{\subsection}{\Large\bfseries}
\titleformat*{\subsubsection}{\large\bfseries}
\titleformat*{\paragraph}{\large\bfseries}
\titleformat*{\subparagraph}{\large\bfseries}
\usepackage{bm}
\usepackage{amsfonts}
\usepackage{latexsym}
\usepackage[latin1]{inputenc}
\usepackage{graphicx}
\usepackage{amsmath}
\usepackage{palatino}
\usepackage{mathpazo}
\usepackage{textcomp}
\usepackage{float}
\usepackage{booktabs}
\usepackage{dcolumn}
\usepackage{ragged2e}
\usepackage{hyperref}
\hypersetup{colorlinks,citecolor=blue}
\hypersetup{colorlinks=true,linkcolor=blue,filecolor=magenta,urlcolor=blue}
\usepackage{amsthm}
\usepackage{xcolor}
\usepackage{orcidlink}
\usepackage{epsfig}
\usepackage{commath}
\usepackage{cancel, ulem}
\usepackage{subcaption}
\usepackage{caption}
\usepackage{multirow}
\newcommand{\m}{\mathring}
\newcommand{\be}{\begin{equation}}
\newcommand{\ee}{\end{equation}}
\newcommand{\bea}{\begin{eqnarray}}
\newcommand{\eea}{\end{eqnarray}}
\newcommand{\eeas}{\end{eqnarray*}}
\newcommand{\beas}{\begin{eqnarray*}}

\def\jnl@style{\it}
\def\aaref@jnl#1{{\jnl@style#1}}

\def\aaref@jnl#1{{\jnl@style#1}}

\def\aj{\aaref@jnl{AJ}}                   
\def\apj{\aaref@jnl{ApJ}}                 
\def\apjl{\aaref@jnl{ApJ}}                
\def\apjs{\aaref@jnl{ApJS}}               
\def\apss{\aaref@jnl{Ap\&SS}}             
\def\aap{\aaref@jnl{A\&A}}                
\def\aapr{\aaref@jnl{A\&A~Rev.}}          
\def\aaps{\aaref@jnl{A\&AS}}              
\def\mnras{\aaref@jnl{Mon.~Not.~Roy.~Astron.~Soc.}}             
\def\prd{\aaref@jnl{Phys.~Rev.~D}}        
\def\prc{\aaref@jnl{Phys.~Rev.~C}}  
\def\prl{\aaref@jnl{Phys.~Rev.~Lett.}}    
\def\qjras{\aaref@jnl{QJRAS}}             
\def\skytel{\aaref@jnl{S\&T}}             
\def\ssr{\aaref@jnl{Space~Sci.~Rev.}}     
\def\zap{\aaref@jnl{ZAp}}                 
\def\nat{\aaref@jnl{Nature}}              
\def\aplett{\aaref@jnl{Astrophys.~Lett.}} 
\def\apspr{\aaref@jnl{Astrophys.~Space~Phys.~Res.}} 
\def\physrep{\aaref@jnl{Phys.~Rep.}}      
\def\physscr{\aaref@jnl{Phys.~Scr}}       
\def\commat{\aaref@jnl{Comm.~Math.~Phys.}}              
\def\science{\aaref@jnl{Science}}               
\def\cqg{\aaref@jnl{Classical Quant.~Grav.}}            
\def\jpcs{\aaref@jnl{JPCS}}                                     
\def\ijmpd{\aaref@jnl{Int.~J.~Mod.~Phys.~D}}                    
\def\grg{\aaref@jnl{Gen.~Relat.~Gravit.}}               
\def\rpp{\aaref@jnl{Rep.~Prog.~Phys.}}          
\def\npa{\aaref@jnl{Nucl.~Phys.~A}}        
\def\lrr{\aaref@jnl{Living Rev.~Rel.}}                   
\def\jcap{\aaref@jnl{J.~Cosmology Astropart.~Phys.}}    
\def\rmp{\aaref@jnl{Rev.~Mod.~Phys.}}   
\def\epjc{\aaref@jnl{Eur.~Phys.~J.~C}} 
\def\plb{\aaref@jnl{~Phy.~Lett.~B}} 
\def\mpla{\aaref@jnl{Mod.~Phy.~Lett.~A}} 
\def\arxiv{\aaref@jnl{arxiv.org}}

\allowdisplaybreaks[1]

\begin{document}

\title{Can an Extra Degree of Freedom in Scalar-Tensor Non-Metricity Gravity Account for the Evolution of the Universe?}
\author{Ghulam Murtaza\orcidlink{0009-0002-6086-7346}}
\email{ghulammurtaza@1utar.my}
\affiliation{Department of Mathematical and Actuarial Sciences, Universiti Tunku Abdul Rahman, Jalan Sungai Long,
43000 Cheras, Malaysia}
\author{Avik De\orcidlink{0000-0001-6475-3085}}
\email{avikde@um.edu.my}
\affiliation{Institute of Mathematical Sciences, Faculty of Science, Universiti Malaya, 50603 Kuala Lumpur, Malaysia}
\author{Andronikos Paliathanasis\orcidlink{0000-0002-9966-5517}}
\email{anpaliat@phys.uoa.gr}
\affiliation{Department of Mathematics, Faculty of Applied
Sciences, Durban University of Technology, Durban 4000, South Africa}
\affiliation{Departamento de Matem\`{a}ticas, Universidad Cat\`{o}lica del 
Norte, Avda.
Angamos 0610, Casilla 1280 Antofagasta, Chile}
\affiliation{School for Data Science and Computational Thinking, Stellenbosch 
University,
44 Banghoek Rd, Stellenbosch 7600, South Africa}
\affiliation{National Institute for Theoretical and Computational Sciences (NITheCS), South Africa.}
\author{Tee-How Loo\orcidlink{0000-0003-4099-9843}}
\email{looth@um.edu.my}
\affiliation{Institute of Mathematical Sciences, Faculty of Science, Universiti Malaya, 50603 Kuala Lumpur, Malaysia}


\footnotetext{The research has been carried out under Universiti Tunku Abdul Rahman Research Fund project IPSR/RMC/UTARRF/2023-C1/A09 provided by Universiti Tunku Abdul Rahman. AP thanks the support of VRIDT through Resoluci\'{o}n VRIDT No. 096/2022 and Resoluci\'{o}n VRIDT No. 098/2022. AP was Financially supported by FONDECYT 1240514 ETAPA 2025. }

\begin{abstract}
 We investigate whether the extra scalar degree of freedom that arises in the second connection class of scalar-tensor non-metricity gravity can accurately replicate and potentially enrich the cosmic expansion history. Focusing on a spatially flat FLRW background, we introduce Hubble-normalized variables and recast the field equations into an autonomous dynamical system. Four representative scenarios are analyzed comprehensively.  Phase-space research reveals a rich hierarchy of critical points: matter-dominated, stiff-fluid, and de Sitter solutions, together with asymptotic trajectories leading to Big-Crunch/Rip singularities and transient, unstable matter epochs. With suitable parameter choices, the standard $\Lambda$CDM sequence is reinstated; however, novel late-time and high-curvature regimes arise exclusively from the non-metricity sector. A systematic comparison of metric scalar-tensor and teleparallel scalar-torsion theories reveals unique stability characteristics and potential observational discriminants. Our findings indicate that the additional time-dependent function inherent to scalar-tensor non-metricity gravity can effectively explain the Universe's evolution while providing new phenomenology that can be tested by upcoming surveys.
\end{abstract}

\maketitle

\tableofcontents
\section{Introduction}\label{sec00}

The mystery of the dark energy, which is supposed to be responsible for the current accelerated expansion of the universe, is still unsettled, especially as the standard model of $\Lambda$CDM faces severe tensions with cosmological data \cite{Perivolaropoulos2022,white,Carloni:2024zpl}, forces to think beyond the theory of general relativity (GR) \cite{Heisenberg2019, Ferrari:2025egk,Wang:2025bkk,Ye:2025ark,Yang:2025mws,Alfano:2024fzv,Paliathanasis:2025dcr,Cai:2025mas,Silva:2025hxw,Duchaniya:2025oeh}. One of the simplest and often investigated scalar-tensor extension of GR is achieved by introducing a scalar field non-minimally coupled to the standard tensor degree of freedom \cite{Bezrukov2008, Akrami2020,Perrotta1999, Boisseau2000,Ballardinii2020}. On the other hand, the dynamics of GR can be derived through approaches other than the Einstein-Hilbert Lagrangian involving the Levi-Civita curvature scalar $\mathring{R}$ \cite{Beltran2019}. It is possible to achieve a teleparallel framework, defined by a zero curvature tensor, by incorporating torsion or non-metricity into the connection. In the first scenario, a flat spacetime with torsion and a metric-compatible affine connection replaces the standard torsion-free, metric-compatible Levi-Civita connection used in GR. Einstein himself introduced this theory \cite{1}, which is referred to as the metric teleparallel theory. In the latter case, the symmetric teleparallel theory arises, built upon an affine connection with zero curvature and torsion \cite{Nester}. In these theories, torsion scalar $\mathbb{T}$ and non-metricity scalar  $Q$ are built respectively from the torsion tensor $T^{\mu}_{\,\,\,\nu\sigma}$ and non-metricity tensor $Q_{\mu\nu\sigma}$. By adopting $\mathcal{L}=\sqrt{-g} \, \mathbb{T}$ for the former and $\mathcal{L}=\sqrt{-g}Q$ for the latter, one derives the field equations.

Since both metric and symmetric teleparallel theories are equivalent to GR, they naturally share its challenges regarding the dark sector, requiring either extensions to the Standard Model of particle physics or the presence of an unknown negative energy component to account for cosmic acceleration. 
To address this problem, modified gravity theories such as $f(\mathbb{T})$ \cite{fT1} and $f(Q)$ \cite{coincident,Heisenberg:2023lru} have been proposed, analogous to how 
$f(\m R)$ gravity extends GR \cite{fR}.
These extensions stand apart from their curvature-based GR counterparts, making them particularly intriguing for exploring novel physical phenomena. Affine connection components are independent from the metric tensor in these theories, further introducing intricate geometries in the formulations. In spatially flat, homogeneous, and isotropic FLRW cosmology, metric teleparallelism allows limited connection families without new functions \cite{Hohmann20199, Hohmann2021}, while symmetric teleparallelism permits three families, each involving an extra time-dependent function, leading to distinct field equations and dynamics \cite{Hohmannnn2021, fQfT1}.

While $f(Q)$ theories offer an interesting runway to take off in terms of gravity research \cite{Carloni:2025kev,Koussour:2024wtt,Narawade:2024pxb,Jensko:2024bee,Alwan:2024lng,Yadav:2024vmt,Ayuso:2021vtj,DAmbrosio:2021zpm,Ferreira:2023awf,Dimakis:2025jrl,Dialektopoulos:2025ihc,Ayuso:2025vkc}, completely uncharted territory has been introduced in \cite{intro1} by non-minimally coupling a scalar field $\phi$ to the non-metricity scalar $Q$ \cite{and01}. The motivation is clearly drawn from the scalar-tensor extension of GR, in which the Ricci scalar was coupled to the scalar field \cite{scalartensor,quiros}, and the scalar-torsion theory, in which the torsion scalar was coupled to the scalar field \cite{saridakisST,phiT}. The cosmological stability aspects of this new theory corresponding to the three possible connection classes were studied, and it was shown that the time-dependent function decouples completely from the cosmological field equations in case of the first connection, reducing the theory to the familiar scalar-torsion model in metric teleparallelism \cite{pati}. The remaining two classes retain the time-dependent function within their dynamical equations, indicating the presence of an extra degree of freedom. In this work, we analyze the contribution of this time-dependent arbitrary function by the cosmological dynamical system analysis technique. 

In the context of cosmology, the field equations are typically presented as a system of coupled, non-linear ordinary differential equations. Since non-linear differential equations lack general solution methods, dynamical system analysis (DSA) provides a powerful method for understanding their qualitative behavior. To construct a dynamical system, one first introduces appropriate dimensionless normalized variables that transform the field equations into a closed system of coupled first-order non-linear differential equations. The universe's evolution through various epochs can be traced using trajectories in phase space. Cosmological epochs like inflation, matter/radiation dominance, and dark energy-driven expansion are often reflected as stationery points in the phase space. To gain insights into the evolution of the universe and its behavior in cosmological models, DSA gained prominence as a valuable approach \cite{Bahamonde2018b, Wainwright1997,Coley2013,Das2019,Grandaa2018,DAgostino:2021vvv,Bargach2019, Mandal2023, Dimakis2024, Matsumoto2018, Hrycyna2015, Cid2018,Odintsov2017a,Roy2024, Chaterje2024,Louw2024,Carloni:2024ybx}. In $f(Q)$ gravity, a series of interesting DSA works was done \cite{Paliathanasis:2023raj,73,74,Channuie2024,solanki2024,narawade2023,4,fQBI,shabani2024a}. Utilizing three different classes of affine connections on a FLRW universe, the DSA in $f(Q)$ was carried out in \cite{Shabani2023a,shabani2023b}. The complete phase space of $f(Q)$-cosmology for all the families of connections investigated in \cite{andfq,Paliathanasis:2023raj}. The boundary term's contribution in non-metricity gravity was examined using DSA techniques recently in \cite{Paliathanasis:2024yea,Lohakare2024, Murtaza2025,shabanifQC}. Very recently, the phase space evolution was explored in the currently studied scalar non-metricity theory with a Chameleon mechanism \cite{Paliathanasiss2024}.

In the present article, we conduct the dynamical system analysis using the Hubble normalization method to study the dynamics and asymptotic solutions of scalar-tensor theory in the non-metricity approach, specifically for the second connection class. This allows us to rewrite the field equations as an equivalent system of algebraic differential equations. We identify the fixed points and examine the physical characteristics of the asymptotic solutions associated with them.

The present article is organized as follows: After the Introduction in Section \ref{sec00}, we provide a brief overview of the mathematical foundations of symmetric teleparallel theory, followed by the field equations for the non-metricity approach of the scalar-tensor gravity in Section \ref{sec1}. Section \ref{sec3} discusses the energy conservation law, while Section \ref{sec4} explores the cosmological implications of this theory and the three varied formulations on a spatialy flat FLRW universe. A comprehensive dynamical systems analysis of the scalar-tensor theory for the second connection class, considering various choices of the coupling function and potential, is presented in Section \ref{sec6} and its subsections. A brief comparison of our result with the scalar-tensor extension of GR and metric teleparallel theory can be found in Section \ref{sec8}, this is beneficial to gauge the role of the extra degrees of freedom available in our study. Finally, our main results and conclusions are summarized in Section \ref{sec7}.

\section{Non-metricity version of scalar-tensor gravity theory} \label{sec1}
The Levi-Civita connection $\mathring{\Gamma}^\alpha{}_{\mu\nu}$ is the unique affine connection with the combined property of metric-compatibility and torsion-free and thus it can be presented in terms of the metric $g$
\begin{equation}
\mathring{\Gamma}^\alpha_{\,\,\,\mu\nu}=\frac{1}{2}g^{\alpha\beta}\left(\partial_\nu g_{\beta\mu}+\partial_\mu g_{\beta\nu}-\partial_\beta g_{\mu\nu}  \right)\,,
\end{equation}
However, we can always consider a torsion-free and curvature-free affine connection $\Gamma^\alpha{}_{\mu\nu}$, with the property of non-vanishing non-metricity tensor
\begin{equation} \label{Q tensor}
Q_{\lambda\mu\nu} := \nabla_\lambda g_{\mu\nu}=\partial_\lambda g_{\mu\nu}-\Gamma^{\beta}_{\,\,\,\lambda\mu}g_{\beta\nu}-\Gamma^{\beta}_{\,\,\,\lambda\nu}g_{\beta\mu}\neq 0 \,.
\end{equation}
We present
\begin{equation} \label{connc}
\Gamma^\lambda{}_{\mu\nu} := \mathring{\Gamma}^\lambda{}_{\mu\nu}+L^\lambda{}_{\mu\nu}
\end{equation}
where $L^\lambda{}_{\mu\nu}$ is the disformation tensor, given by
\begin{equation} \label{L}
L^\lambda{}_{\mu\nu} = \frac{1}{2} (Q^\lambda{}_{\mu\nu} - Q_\mu{}^\lambda{}_\nu - Q_\nu{}^\lambda{}_\mu) \,.
\end{equation}

The superpotential (or the non-metricity conjugate) tensor $P^\lambda{}_{\mu\nu}$ is given by
\begin{equation} \label{P}
P^\lambda{}_{\mu\nu} = 
\frac{1}{4} \left( -2 L^\lambda{}_{\mu\nu} + Q^\lambda g_{\mu\nu} - \tilde{Q}^\lambda g_{\mu\nu} -\delta^\lambda{}_{(\mu} Q_{\nu)} \right) \,,
\end{equation}
where
\begin{equation*}
 Q_\mu := g^{\nu\lambda}Q_{\mu\nu\lambda} = Q_\mu{}^\nu{}_\nu \,, \qquad \tilde{Q}_\mu := g^{\nu\lambda}Q_{\nu\mu\lambda} = Q_{\nu\mu}{}^\nu \,.
\end{equation*}
Finally, the non-metricity scalar $Q$ is defined as
\begin{equation} \label{Q}
Q=Q_{\alpha\beta\gamma}P^{\alpha\beta\gamma}\,.
\end{equation}

Mimicking the gravitational action of the scalar-tensor extension of GR, an action was considered in \cite{intro1}
\begin{align}\label{eqn:ST}
S=\frac{1}{2\kappa }\int\sqrt{-g}\left[f(\phi)Q-h(\phi)\nabla^\alpha\phi\nabla_\alpha\phi-U(\phi)
+2\kappa\mathcal L_m \right] \,d^{4}x\,.
\end{align}
$U(\phi)$ is the scalar field potential, $f(\phi)$ couples the scalar field to the non-metricity scalar $Q$. As well-known in scalar-tensor theory, the action is invariant under scalar field reparametrization, which can reduce one of the functions to be a constant. So, without loss of generality, let us redefine the scalar fields to make $h(\phi)=1$. The variation of the action term with respect to the metric produces the metric field equations
\begin{align}
\kappa T_{\mu\nu}
=&f\m G_{\mu\nu} +
2f'P^\lambda{}_{\mu\nu}\nabla_\lambda \phi
-h\nabla_\mu\Phi\nabla_\nu\phi
+\frac12hg_{\mu\nu}\nabla^\alpha\phi\nabla_\alpha\phi+\frac12Ug_{\mu\nu} \,,
\label{eqn:FE1}
\end{align}
where 
$\m G_{\mu\nu}$ denotes the Einstein tensor corresponding to the Levi-Civita connection;
$T_{\mu\nu}$ is the stress energy tensor defined as 
\begin{align*}
T_{\mu\nu}=-\frac 2{\sqrt{-g}}\frac{\delta(\sqrt{-g}\mathcal L_M)}{\delta g^{\mu\nu}}\,,
\end{align*}
and 
(~)' means the derivative of the function (~) with  respect to $\phi$.
On the other hand, the variation of the action with respect to the scalar field $\phi$ leads us to the second field equations  
\begin{align}\label{eqn:FE2}
f'Q+h'\nabla^\alpha\phi\nabla_\alpha\phi+2h\m\nabla^\alpha\m\nabla_\alpha \phi-U'=0.
\end{align}
Apart from the metric tensor and the scalar field $\phi$, there is another set of dynamic variables: the components of the affine connection, which yields the connection field equations 
\begin{align}\label{eqn:FE3}
(\nabla_\mu-\tilde L_\mu)(\nabla_\nu-\tilde L_\nu)
\left[4fP^{\mu\nu }{}_\lambda+\kappa\Delta_\lambda{}^{\mu\nu}\right]=0\,,
\end{align}
where 
\[\Delta_\lambda{}^{\mu\nu}=-\frac2{\sqrt{-g}}\frac{\delta(\sqrt{-g}\mathcal L_M)}{\delta\Gamma^\lambda{}_{\mu\nu}}\,,\]is the hypermomentum tensor
\cite{hyper}.

The effective stress energy tensor $T^{\text{eff}}_{\mu\nu}$ is constructed using the relation
\begin{align*}
    f\m G_{\mu\nu}=\kappa T^{\text{eff}}_{\mu\nu}\,,
\end{align*}
where
\begin{equation} \label{T^eff}
 T^{\text{eff}}_{\mu\nu} 
 =  T_{\mu\nu}+ \frac 1{\kappa}\left[
        -2f'P^\lambda{}_{\mu\nu}\nabla_\lambda \phi
        +h\nabla_\mu\phi\nabla_\nu\phi 
        -\frac12hg_{\mu\nu}\nabla^\alpha\phi\nabla_\alpha\phi
        -\frac12 Ug_{\mu\nu}\right]\,.
\end{equation}
The additional part in (\ref{T^eff}) describes a source of fictitious dark energy that can drive the late-time acceleration driven by a negative pressure 
\begin{align}
T^{\text{DE}}_{\mu\nu}= \frac 1{f}\left[
        -2f'P^\lambda{}_{\mu\nu}\nabla_\lambda \phi
        +h\nabla_\mu\phi\nabla_\nu\phi 
        -\frac12hg_{\mu\nu}\nabla^\alpha\phi\nabla_\alpha\phi
        -\frac12 Ug_{\mu\nu}\right]\,.
\end{align}

In the present paper, we consider a perfect fluid type stress energy tensor given by
\begin{align}
T_{\mu\nu}=pg_{\mu\nu}+(p+\rho)u_\mu u_\nu
\end{align}
where $\rho$, $p$ and $u^\mu$ denote 
the energy density, pressure, and four velocity of the fluid, respectively.


\section{Energy conservation - a note}\label{sec3}
Unlike the scalar-tensor extension of GR, the non-metricity approach is not straightaway compatible with the classical energy conservation condition. 
The divergence of  (\ref{eqn:FE1}) yields  
\begin{align}\label{vv0}
\kappa\m\nabla_\mu T^\mu{}_\nu
=&\left(\m G^\lambda{}_\nu+2\m\nabla_\mu P^{\lambda\mu}{}_\nu\right)
\nabla_\lambda f
     +2P^{\lambda\mu}{}_\nu\m\nabla_\mu\nabla_\lambda f
     +\left(-h\m\nabla^\alpha\nabla_\alpha\phi 
        -\frac12h'\m\nabla^\alpha\m\nabla_\alpha\phi+\frac12U'\right)\phi_\nu\,.
\end{align}

In view of (\ref{eqn:FE2}), the equation (\ref{vv0}) becomes 
\begin{align}
\kappa\m\nabla_\mu T^\mu{}_\nu
 =&\left(\m G^\lambda{}_\nu+\frac Q2\delta^\lambda{}_\nu\right) \nabla_\lambda f 
 +2\m\nabla_\mu\left(P^{\lambda\mu}{}_\nu\nabla_\lambda f\right)\,.
\end{align}
Using the following result derived in \cite{fQC},
\begin{align}
2(\nabla_\lambda&-\tilde L_\lambda)(\nabla_\mu-\tilde L_\mu)
 (f P^{\lambda\mu}{}_\nu) 
 =\left(\m G^\lambda{}_\nu+\frac Q2\delta^\lambda{}_\nu\right) \nabla_\lambda f 
 +2\m\nabla_\mu\left(P^{\lambda\mu}{}_\nu\nabla_\lambda f\right)\,,
 \label{eqn:C1a}
\end{align}
we can finally conclude that

\begin{align}\label{eqn:EC}
\kappa\m\nabla_\mu T^\mu{}_\nu
=&2(\nabla_\lambda-\tilde L_\lambda)(\nabla_\mu-\tilde L_\mu)
 (f P^{\lambda\mu}{}_\nu) \notag\\
\end{align}
Hence, we can conclude that energy conservation is fulfilled provided the matter Lagrangian is independent of an affine connection, in view of (\ref{eqn:FE3}).
\section{Cosmological application}\label{sec4}
We consider a homogeneous and isotropic Friedmann-Lema\^{i}tre-Robertson-Walker (FLRW) spacetime given by 
\begin{align}\label{ds:RW}
ds^2=-dt^2+a^2\left(\frac{dr^2}{1-kr^2}+r^2d\theta^2+r^2\sin^2\theta d\phi^2\right)
\end{align}
with scale factor $a(t)$, Hubble parameter $H=\dot a/a$ and the spatial curvature $k=0,+1,-1$ respectively 
modeled the universe of spatially flat, closed, and open types. In the present work, we study a spatially flat spacetime.
Here the $\dot{(~)}$ denotes the derivative with respect to $t$.
In addition, we denote $u_\mu=(dt)_\mu$ and  
$h_{\mu\nu}=g_{\mu\nu}+u_\mu u_\nu$.

There are three classes of affine connections that are compatible with the symmetric teleparallel framework, which are given as follows: \cite{fQfT1, FLRW/connection} 
\begin{align} \label{eqn:conn}
\Gamma^t{}_{tt}=&C_1, 
	\quad 					\Gamma^t{}_{rr}=\frac{C_2}{\chi^2}, 
	\quad 					\Gamma^t{}_{\theta\theta}=C_2r^2, 
	\quad						\Gamma^t{}_{\phi\phi}=C_2r^2\sin^2\theta,								\notag\\
\Gamma^r{}_{tr}=&C_3, 
	\quad  	\Gamma^r{}_{rr}=\frac{kr}{\chi^2}, 
	\quad		\Gamma^r{}_{\theta\theta}=-\chi^2r, 
	\quad		\Gamma^r{}_{\phi\phi}=-\chi^2r\sin^2\theta,												\notag\\
\Gamma^\theta{}_{t\theta}=&C_3, 
	\quad		\Gamma^\theta{}_{r\theta}=\frac1r,
	\quad		\Gamma^\theta{}_{\phi\phi}=-\cos\theta\sin\theta,										\notag\\
\Gamma^\phi{}_{t\phi}=&C_3, 
	\quad 	\Gamma^\phi{}_{r\phi}=\frac1r, 
	\quad 	\Gamma^\phi{}_{\theta\phi}=\cot\theta,
\end{align}
where $C_1$, $C_2$, and $C_3$ are temporal functions, which must fulfill one of the following criteria:  
\begin{enumerate}\label{casesConn}
\item[(I)] $C_1=\gamma$, $C_2=C_3=0$ and $k=0$, where $\gamma$ is a temporal function; or
\item[(II)] $C_1=\gamma+\dfrac{\dot\gamma}\gamma$, $C_2=0$, $C_3=\gamma$ and $k=0$,
             where $\gamma$ is a nonvanishing temporal function; or 
\item[(III)] $C_1=-\dfrac k\gamma-\dfrac{\dot\gamma}{\gamma}$, $C_2=\gamma$, $C_3=-\dfrac k\gamma$ and $k=0,\pm1$,
             where $\gamma$ is a nonvanishing temporal function.
\end{enumerate} 
Hence within the formulation (\ref{eqn:conn}),
the disformation tensor; the superpotential tensor, and the non-metricity scalar 
can be derived respectively as 
\begin{align}
L^\lambda{}_{\mu\nu}
=&C_1u^\lambda u_\mu u_\nu
+\left(\frac{C_2}{a^2}-H\right)u^\lambda h_{\mu\nu}
+(C_3-H)(-h^\lambda{}_\nu u_\mu-h^\lambda{}_\mu u_\nu)  \label{eqn:L2}\\
2P^\lambda{}_{\mu\nu}
=&\frac12\left(3C_3-3\frac{C_2}{a^2}\right)u^\lambda u_\mu u_\nu
    +\frac12\left(3C_3+\frac{C_2}{a^2}-4H\right)u^\lambda h_{\mu\nu}
    +\frac12(C_1+C_3-H)(-h^\lambda{}_\mu u_\nu-h^\lambda{}_\nu u_\mu) 
    \label{eqn:P2}\\
Q
=&3\left(-2H^2+2\frac k{a^2}+3HC_3+\dot C_3+\frac1{a^2}(HC_2+\dot C_2)\right)\,.
\label{eqn:Q2}
\end{align}
Hence, the Friedmann-like equations can be derived
\begin{align}
\kappa p
=&f\left(-2\dot H-3H^2-\frac k{a^2}\right)
    +\frac12\dot f\left(3C_3+\frac{C_2}{a^2}-4H\right)
    -\frac12h\dot \phi^2    +\frac12U\,,\\
\kappa\rho
=&f\left(3H^2+3\dfrac k{a^2}\right)
    +\frac12\dot f\left(3C_3-3\frac{C_2}{a^2}\right)
     -\frac12h\dot \phi^2    -\frac12U\,,\\
0=&\partial_j\phi=\partial_j f
    ; \quad j=1,2,3\,.
\end{align}

\begin{align}
p^{\text{DE}}
=&-\frac1{2f}\left[ 
    \dot f\left(3C_3+\frac{C_2}{a^2}-4H\right)
    -h\dot \phi^2    +U  \right]\,,\\
\rho^{\text{DE}}
=&-\frac1{2f}\left[
    \dot f\left(3C_3-3\frac{C_2}{a^2}\right)
     -h\dot \phi^2    -U \right]\,.
\end{align}
Moreover, the modified continuity relation can be directly deduced from 
(\ref{eqn:EC}) 
\begin{align}
\kappa\dot{\rho}+3\kappa H(\rho+p)
=\frac32\left[
    \dot f\left(3C_3H-\frac{2\dot C_2+C_2H}{a^2}\right)
    +\ddot f\left(C_3-\frac{C_2}{a^2}\right)\right].
\end{align}


\section{Cosmological dynamical system}\label{sec6}
It is clear that the time-dependent function $\gamma(t)$ does not appear in the cosmological dynamics of connection class I. It is also obvious that the cosmological equations for connection I exactly match the corresponding equations in the scalar-torsion counterpart. In this work we consider the cosmological dynamics of the class of affine connections of type II to investigate the contribution of $\gamma(t)$.  

Recently, it was found that within $f(Q)$-gravity, the dynamical degrees of freedom field equations of the second connection are attributed to scalar fields and the theory corresponds to a quintom-like cosmological model \cite{Basilakos:2025olm}. Furthermore, from the observational constraints in \cite{Paliathanasis:2025hjw}, it was found that the cosmological theory related to the second connection challenges $\Lambda$CDM, as it provides a better fit to the data.

The Friedmann equations in this case are given as follows
\begin{align}\label{peq}
\kappa p
=&\left(-2\dot H-3H^2\right)f
    +\frac12\left(3\gamma-4H\right)\dot f
    -\frac12h\dot \phi^2    +\frac12U\,,
\end{align}
\begin{align}\label{rhoeq}
\kappa\rho
=&3H^2f
    +\frac32\gamma\dot f
     -\frac12h\dot \phi^2    -\frac12U\,.
\end{align}
The scalar field eq (\ref{eqn:FE2}) gives
\begin{align}\label{eqn:FE2-2}
\left(-6H^2+3\left\{\dot\gamma+3H\gamma\right\}\right)f'
-h'\dot \phi^2 -2h(\ddot\phi+3H\dot \phi)-U'=0.
\end{align}
The connection eq is given by
\begin{align} \label{connection:eq}
3\gamma \left(3\dot f H +\ddot f\right)=0.
\end{align}
The continuity relation is given below, and the right side vanishes due to (\ref{connection:eq})
\begin{align}
\kappa\dot{\rho}+3\kappa H(\rho+p)
=\frac32\gamma\left[  3\dot f H +\ddot f \right].
\end{align}
We observe that the extra degree of freedom enters the system only through terms involving derivatives of the coupling function $f(\phi)$, rendering it irrelevant unless a non-minimal coupling is present. 

From eq (\ref{rhoeq}), we have
\begin{align}
\frac{\kappa \rho}{3fH^2}=1+\frac{\gamma \dot{\phi}f'}{2fH^2}-\frac{h \dot{\phi}^2}{6fH^2}-\frac{U}{6fH^2},
\end{align}
In the following, $\overline{(.)}$ represents derivative with respect to $N=lna$ or $d/dN$. Also, we consider a pressurless dust era in this study. We define the variables:\\
\begin{align}\label{connIIvar}
x^2=\frac{\dot{\phi^2}}{6fH^2},~~y=\frac{U}{3fH^2},~~z=\frac{\sqrt{3}\gamma f'}{\sqrt{2f}H}, ~~s=\frac{\kappa \rho}{3fH^2}, ~~ \lambda=\frac{U' \sqrt{f}}{U}, ~~ \zeta =\frac{f'}{\sqrt{f}},~~\Gamma=\frac{U'' U}{U'^2},~~\Delta=\frac{f''f}{f'^2},
\end{align}
The constraint equation is
\begin{align} {\label{constr}}
 s=1+xz-hx^2 -y   .
\end{align}
From eq (\ref{peq}) by considering $p=0$, we can get following
\begin{align} {\label{dH}}
 \frac{\dot{H}}{H^2}=-\frac{3}{2}+\frac{3xz}{2}-\sqrt{6}x\zeta-\frac{3hx^2}{2}+\frac{3y}{4}   ,
\end{align}
 Utilizing (\ref{peq}), (\ref{eqn:FE2-2}) and (\ref{connection:eq}), the general dynamical system equations can be written as,
\begin{equation}
\overline{x}=-\frac{3x}{2}-\frac{3x^2 z}{2}+\frac{3hx^3}{2}-\frac{3xy}{4}-\frac{x^2 \sqrt{6f}f''}{f'}+\frac{x^2\sqrt{6}f'}{2\sqrt{f}},
\end{equation}

\begin{equation}
\overline{y}=\sqrt{6}\lambda xy+3y-3yzx+3hx^2 y-\frac{3y^2}{2}+\frac{\sqrt{6}xyf'}{\sqrt{f}},
\end{equation}
\begin{equation}
    \overline{z}=\frac{\sqrt{6}f'}{\sqrt{f}}-\frac{3z}{2}-\frac{2\sqrt{6f}f''x^2h}{f'}+\frac{\sqrt{3}\lambda y}{\sqrt{2}}+\frac{\sqrt{6f}f''xz}{f'}-\frac{3z^2 x}{2}+\frac{\sqrt{3}f'xz}{\sqrt{2f}}+\frac{3hx^2z}{2}-\frac{3yz}{4},
\end{equation}
\begin{align}
\overline{s}=2s(1+\omega)+\frac{\sqrt{6}xsf'}{\sqrt{f}}-3sxz-3s+3hx^2s-\frac{3}{2}ys,
\end{align}
\begin{align}
 \overline{\lambda}=\frac{\sqrt{6}xfU''}{U}+\frac{\sqrt{3}x \lambda f'}{\sqrt{2}\sqrt{f}}-\sqrt{6}x\lambda^2,
\end{align}
\begin{align}
 \overline{\zeta}=\sqrt{6}xf''-\frac{\sqrt{3}xf'^2}{\sqrt{2}f}.
\end{align}
Now using the constraint $(\ref{constr})$, we can eliminate the variable $s$, so our autonomous dynamical system can be rewritten as

\begin{equation}
\overline{x}=-\frac{3x}{2}-\frac{3x^2 z}{2}+\frac{3hx^3}{2}-\frac{3xy}{4}- \sqrt{6}x^2 \Delta \zeta+\frac{x^2\sqrt{6}\zeta}{2},
\end{equation}
\begin{equation}
\overline{y}=\sqrt{6}\lambda xy+3y-3yzx+3hx^2 y-\frac{3y^2}{2}+\sqrt{6}xy\zeta,
\end{equation}
\begin{equation}
    \overline{z}=\sqrt{6}\zeta-\frac{3z}{2}-2\sqrt{6}x^2h\Delta \zeta+\frac{\sqrt{3}\lambda y}{\sqrt{2}}+\sqrt{6}xz \Delta \zeta-\frac{3z^2 x}{2}+\frac{\sqrt{3}xz\zeta}{\sqrt{2}}+\frac{3hx^2z}{2}-\frac{3yz}{4},
\end{equation}
\begin{align}{\label{lambda:new}}
 \overline{\lambda}=\sqrt{6}x\lambda^2(\Gamma-1)+\frac{\sqrt{3}x\lambda \zeta}{\sqrt{2}},
\end{align}
\begin{align}\label{zeta:eq}
 \overline{\zeta}=\sqrt{6}x\zeta^2(\Delta -\frac{1}{2}).
\end{align}
Finally, the equation of state parameter for total fluid i.e $w_{eff}=-1-\frac{2}{3}\frac{\dot{H}}{H^2}$, is expressed as
\begin{align}
    w_{eff}=-xz+\frac{2\sqrt{2}x \zeta}{\sqrt{3}}+hx^2 -\frac{y}{2},
\end{align}
The deceleration parameter $q=-1-\frac{\dot{H}}{H^2}$, is given as
\begin{align}
   q=\frac{1}{2}-\frac{3xz}{2}+\sqrt{6}x\zeta +\frac{3hx^2}{2}-\frac{3y}{4}. 
\end{align}
Here, we divide the scenario into four distinct cases. One can notice that we have only two functions for the scalar field i.e., $\zeta$ and $\lambda$.

\subsection{$\lambda=\lambda_0$, $\zeta=\zeta_0$}\label{case I}
In such case, from (\ref{connIIvar}) we can determine the coupling function $f(\phi)=\frac{\zeta_0^2 \phi^2}{4}$ and the potential $U=U_{0} \phi^{\frac{2 \lambda_0}{\zeta_0}}$, where $U_{0}$ and $\zeta_0$ are constant.
\footnote{Let us give a clarification here regarding a generalized coupling $f(\phi)$. The fixed points satisfy (\ref{zeta:eq}), 
$$\sqrt{6}x\zeta^2(\Delta(f) -\frac{1}{2})=0$$
which is satisfied either when $x=0$ or for $\zeta=\zeta_0$, such that $\zeta_0(\Delta(f)-\frac12)=0$. 

In the first case, when $x=0$, $\lambda$ is not constrained at all by the system.

The latter case yields an exponential or power law coupling function as evident from (\ref{connIIvar}). Subsequently, the remaining equations impose constraints on the variables $x, y, z$, yielding families of fixed points determined by the structure of the coupling function. Consequently, the formulation of an arbitrary coupling function will not result in the development of novel families of solutions. Therefore, the examination of the exponential and power-law coupling functions is adequate to derive all potential solutions. Nonetheless, it is crucial to note that the stability characteristics of the equilibrium points will not remain unchanged. and are dependent on the function $\Delta$.} So, our dynamical system reduces to 3D, and we can write it in an autonomous way as follows
\begin{equation}\label{eq1}
\overline{x}=-\frac{3x}{2}-\frac{3x^2 z}{2}+\frac{3hx^3}{2}-\frac{3xy}{4},
\end{equation}
\begin{equation}\label{eq2}
\overline{y}=\sqrt{6}\lambda_0 xy+3y-3yzx+3hx^2 y-\frac{3y^2}{2}+\sqrt{6}xy\zeta_0,
\end{equation}
\begin{equation}\label{eq3}
    \overline{z}=\sqrt{6}\zeta_0-\frac{3z}{2}-\sqrt{6}x^2h \zeta_0+\frac{\sqrt{3}\lambda_0 y}{\sqrt{2}}-\frac{3z^2 x}{2}+\sqrt{6}xz\zeta_0+\frac{3hx^2z}{2}-\frac{3yz}{4}.
\end{equation}

\begin{table}[h]
    \centering
    \begin{tabular}{|c|c|c|c|c|c|}
        \hline
        Critical point  &$(x,y,z)$ &Existence &$w_{eff}$ & $q$& Can be Attractor? \\ \hline
        A & $\left( 0,0,2\sqrt{\frac{2}{3}} \zeta_0 \right)$ &Always&0 & $\frac{1}{2}$& No\\ \hline
        B & $\left( 0,2,\frac{\sqrt{6}}{3} (\zeta_0+\lambda_0) \right)$& Always & $-1$&-1 & Yes\\ \hline
        C & $\left( -\frac{\sqrt{6}}{\zeta_0+\lambda_0},0,\frac{-6h+\zeta_0^2+2\zeta_0 \lambda_0+\lambda_0^2}{\sqrt{6}(\zeta_0+\lambda_0)} \right)$ & $\zeta_0+\lambda_0 \neq 0$&$1-\frac{4\zeta_0 }{\zeta_+\lambda_0}$&$2-\frac{6\zeta_0}{\zeta_0+\lambda_0}$&No\\ \hline
    \end{tabular}
    \caption{Critical points and their physical properties}
    \label{case1table1}
\end{table}
The details of the critical points (CP) for the dynamical system (\ref{eq1})-(\ref{eq3}) are provided in Table \ref{case1table1}, including their existence conditions, attractor nature, equation of state parameter $w_{eff}$, and deceleration parameter $q$.  

The critical point A represents a decelerating dust-dominated universe. The eigenvalues of the linearized system (\ref{eq1})-(\ref{eq3}) around the stationary point A are $e_{1}(A)=3, e_{2}(A)=-\frac{3}{2}$ and $e_{3}(A)=-\frac{3}{2}$, from which we can conclude that A is saddle. 

The stationary point B always exists and the eigenvalues corresponding to this are $e_{1}(B)=-3, e_{2}(B)=-3$ and $e_{3}(B)=-3$, implies the stable nature of this fixed point and it describes the de Sitter universe with $w_{eff}=-1$ and $q=-1$. 

The existence condition of critical point C is mentioned in Table \ref{case1table1}. The physical properties of this CP depends on the constant parameters $\zeta_0$ and $\lambda_0$, i.e for vanishing $\zeta_0$, it provides stiff fluid $w_{eff}=1$ and decelerated universe. The eigenvalues associated to this point $e_{1}(C)=0, e_{2}(C)=0$ and $e_{3}(C)=3$, which gives its unstable behavior.

    

\subsection{$\lambda=\lambda_0$ and $\zeta$ is variable}\label{case II}
We can consider an ansatz for either the potential or the coupling function in this case. A power law or exponential law is the most natural choice. We observe that an exponential potential yields a constant coupling function. So we ignore it. On the other hand, a power law potential function is obtained and analyzed in the previous case \ref{case I}. So we analyze the ansatz $f(\phi)=f_0 e^{\alpha \phi}$. From (\ref{connIIvar}) we get $U=exp (\frac{-2\lambda_0}{\sqrt{f_0}\alpha}e^{-\frac{\alpha \phi}{2}})$\\
\subsubsection{Finite fixed point analysis}
The dynamical system equations for this scenario can be given as
\begin{equation}\label{4}
\overline{x}=-\frac{3x}{2}-\frac{3x^2 z}{2}+\frac{3hx^3}{2}-\frac{3xy}{4}-\frac{\sqrt{6}x^2 \zeta}{2},
\end{equation}
\begin{equation}\label{5}
\overline{y}=\sqrt{6}\lambda_{0} xy+3y-3yzx+3hx^2 y-\frac{3y^2}{2}+\sqrt{6}xy\zeta,
\end{equation}
\begin{equation}\label{6}
    \overline{z}=\sqrt{6}\zeta-\frac{3z}{2}-2\sqrt{6}x^2h \zeta+\frac{\sqrt{3}\lambda_{0} y}{\sqrt{2}}-\frac{3z^2 x}{2}+\frac{3\sqrt{3}xz\zeta}{\sqrt{2}}+\frac{3hx^2z}{2}-\frac{3yz}{4},
\end{equation}
\begin{equation}\label{7}
    \overline{\zeta}=\frac{\sqrt{3}x\zeta^2}{\sqrt{2}}.
\end{equation}

\begin{table}[h]
    \centering
    \resizebox{\textwidth}{!}{ 
    \begin{tabular}{|c|c|c|c|c|c|}
        \hline
        Critical point & $(x,y,z,\zeta)$ & Existence &$w_{eff}$ & $q$& Can be Attractor? \\ \hline
        $A$ & $\left( 0,0,z,\frac{1}{2}\sqrt{\frac{3}{2}} z\right)$ & $z=$ arbitrary& $0$ & $\frac{1}{2}$&No  \\ \hline
       $B$ & $\left( 0,2,z,\frac{1}{2} (\sqrt{6}z-2\lambda_0) \right)$ & $z=$ arbitrary& $-1$ & $-1$&Yes \\ \hline
        $C$ & $\left( x,0,\frac{-1+hx^2}{x},0 \right)$ & $x \neq 0$ & $1$ & $2$&No\\ \hline
          $D$ & $\left( -\frac{\sqrt{6}}{\lambda_0},0,\frac{1}{6}(6hx+\sqrt{6}\lambda_0),0\right)$& $ \lambda_0 \neq 0$&$1+\frac{6h}{\lambda_0^2}+\frac{\sqrt{6}hx}{\lambda_0}$& $2+\frac{9h}{\lambda_0^2}+\frac{3\sqrt{\frac{3}{2}}hx}{\lambda_0}$&No \\ \hline
         
    \end{tabular}
    }
    \caption{Critical points and their physical properties}
    \label{table1_case2}
\end{table}

The CPs for the dynamical system (\ref{5})-(\ref{7}) are listed in Table \ref{table1_case2}, that explain their existence conditions, attractor nature, and cosmological parameters $w_{eff}$ and $q$.  

The critical point A has $w_{eff}=0$, $ q=\frac{1}{2}$, and eigenvalues $e_{1}(A)=3, e_{2}(A)=-\frac{3}{2}$, $e_{3}(A)=-\frac{3}{2}$, and $e_{4}(A)=0$. The asymptotic solution at point A describes a universe dominated by the dust fluid. A is always a saddle point, which means that the asymptotic solution is unstable.

At the point B the matrix of the linearized system has the following eigenvalues $e_{1}(B)=-3, e_{2}(B)=-3$, $e_{3}(B)=-3$ and $e_{4}(B)=-3$, also $w_{eff}=-1$ and $ q=-1$, which means that it is stable de Sitter point.

The stationary point C with $w_{eff}=1$ and $q=2$ describes stiff fluid decelerated universe with unstable nature as the eigenvalues at this are $e_{1}(C)=0, e_{2}(C)=0$, $e_{3}(C)=3$ and $e_{4}(C)=6+\sqrt{6}x\lambda_0$.

The physical properties of critical point D depends on the constant parameters $h$ and $\lambda_0$, i.e for vanishing $h$, it properties are the same as that of fixed point C, the eigenvalues corresponding to this are $e_{1}(D)=-3$, $e_{2}(D)=\frac{3h(6+\sqrt{6}x\lambda_0)}{\lambda_0^2}$, $e_{3}(D)=\frac{3(6h+\sqrt{6}hx\lambda_0)}{2\lambda_0^2}$ and $e_{4}(D)=\frac{3(18h+3\sqrt{6}hx\lambda_0+2\lambda_0^2)}{2\lambda_0^2}$, and it  unstable for $(x>0 \land \lambda_0 >0 \land h \ge0)$.
\subsubsection{ Analysis at infinity}
Since our dynamical variables are not constrained, they can take values at infinity.  Hence, in order to study the analysis at infinity for the system (\ref{5})-(\ref{7}), we introduce the Poincare variables as follows
\begin{equation*}
    x=\frac{X}{\rho}, ~~y=\frac{Y}{\rho}, ~~z=\frac{Z}{\rho}.
\end{equation*}
where $\rho= \sqrt{1-X^2-Y^2-Z^2}$.  We define the new independent variable $dt=\sqrt{1-X^2-Y^2-Z^2} dT$. The equation of state parameter in terms of these new variables is rewritten as
\begin{align}\label{weffinfinity}
    w_{eff}=\frac{6hX^2-6XZ+4\sqrt{6} X \zeta \rho-3Y\rho}{6\rho^2},
\end{align}
Also, the deceleration parameter is expressed as
\begin{align}
   q=\frac{6hX^2-6XZ+4\sqrt{6} X \zeta \rho-3Y\rho+2\rho^2}{4\rho^2}. 
\end{align}

At infinity, the dynamical system (\ref{5})-(\ref{7}) becomes
\begin{align}\label{eq1_case1_infinite}
\frac{dX}{dT}= &-\frac{1}{4} X \Bigg( 
6 (1 + h) X^4 - 12 Y^4 
- 2 Y^3 \left( \sqrt{6} \lambda_{0} Z 
+ 3 \sqrt{1 - X^2 - Y^2 - Z^2} \right) - Y (-1 + Z^2) \left( 
2 \sqrt{6} \lambda_{0} Z 
+ 3 \sqrt{1 - X^2 - Y^2 - Z^2} \right) \nonumber \\
& - 2 X^3 \left( 
3 Z 
+ \sqrt{6} \zeta \sqrt{1 - X^2 - Y^2 - Z^2} \right) + Y^2 \left( 
6 - 6 Z^2 
- 4 \sqrt{6} \zeta Z \sqrt{1 - X^2 - Y^2 - Z^2} \right) \nonumber \\
& + 2 (-1 + Z^2) \left( 
-3 + 3 Z^2 
- 2 \sqrt{6} \zeta Z \sqrt{1 - X^2 - Y^2 - Z^2} \right) 
+ X \left( 
(6 - 12 Y^2) Z - 6 Z^3 
+ 2 \sqrt{6} \left( \zeta + 2 \lambda_{0} Y^2 + 2 \zeta Y^2 \right) \right. \nonumber \\
& \left. \times \sqrt{1 - X^2 - Y^2 - Z^2} + 6 \sqrt{6} \zeta Z^2 \sqrt{1 - X^2 - Y^2 - Z^2} \right) - X^2 \left( 
6 (2 + h) + (6 - 12 h) Y^2 - 6 (2 + h) Z^2 
 + 4 \sqrt{6} (1 + 2 h) \right. \nonumber \\
& \left. \times \zeta Z \sqrt{1 - X^2 - Y^2 - Z^2}+ Y \left( 
2 \sqrt{6} \lambda_{0} Z 
+ 3 \sqrt{1 - X^2 - Y^2 - Z^2} \right) \right) 
\Bigg),
\end{align}

\begin{align}
\frac{dY}{dT}=& \frac{1}{4} Y \Bigg(
-6 (1 + h) X^4 + 12 Y^4 
+ 2 Y^3 \left( \sqrt{6} \lambda_{0} Z 
+ 3 \sqrt{1 - X^2 - Y^2 - Z^2} \right)  + 2 X^3 \left( 3 Z 
+ \sqrt{6} \zeta \sqrt{1 - X^2 - Y^2 - Z^2} \right) \nonumber \\
& 
- 2 (-1 + Z^2) \left( 
6 + 3 Z^2 - 2 \sqrt{6} \zeta Z \sqrt{1 - X^2 - Y^2 - Z^2} \right)  + Y^2 \left( 
-24 + 6 Z^2 + 4 \sqrt{6} \zeta Z \sqrt{1 - X^2 - Y^2 - Z^2} \right) \nonumber \\
& + Y \left( 
-2 \sqrt{6} \lambda_{0} Z + 2 \sqrt{6} \lambda_{0} Z^3 
- 6 \sqrt{1 - X^2 - Y^2 - Z^2} 
+ 3 Z^2 \sqrt{1 - X^2 - Y^2 - Z^2} \right) \nonumber \\
& + 2 X \Big( 
6 (-1 + Y^2) Z + 3 Z^3 
- 2 \sqrt{6} (\lambda_{0} + \zeta) (-1 + Y^2) 
\sqrt{1 - X^2 - Y^2 - Z^2}  - 3 \sqrt{6} \zeta Z^2 
\sqrt{1 - X^2 - Y^2 - Z^2} \Big) \nonumber \\
& + X^2 \Big( 
-6 + 12 h + (6 - 12 h) Y^2 - 6 (2 + h) Z^2 
+ 4 \sqrt{6} (1 + 2 h) \zeta Z \sqrt{1 - X^2 - Y^2 - Z^2} 
\nonumber \\
& + Y \left( 
2 \sqrt{6} \lambda_{0} Z + 3 \sqrt{1 - X^2 - Y^2 - Z^2} \right) \Big)
\Bigg),
\end{align}
\begin{align}
\frac{dZ}{dT} = \frac{1}{4} \Big(&
12 Y^4 Z 
+ 2 Y^3 \left( -\sqrt{6} \lambda_{0} + \sqrt{6} \lambda_{0} Z^2 + 3 Z \sqrt{1 - X^2 - Y^2 - Z^2} \right)  + Y (-1 + X^2 + Z^2)\nonumber \\
& \times \left( -2 \sqrt{6} \lambda_{0} + 2 \sqrt{6} \lambda_{0} Z^2 + 3 Z \sqrt{1 - X^2 - Y^2 - Z^2} \right) + 2 \Big( 
-3 Z^5 
- 2 \sqrt{6} (-1 + (1 + 2 h) X^2) \zeta \sqrt{1 - X^2 - Y^2 - Z^2} \nonumber \\
&\quad + Z^4 \left( 3 X + 2 \sqrt{6} \zeta \sqrt{1 - X^2 - Y^2 - Z^2} \right)  - 3 Z^3 \left( -2 + (2 + h) X^2 + \sqrt{6} X \zeta \sqrt{1 - X^2 - Y^2 - Z^2} \right) \nonumber \\
&\quad + Z^2 \left( -3 X + 3 X^3 - 4 \sqrt{6} \zeta \sqrt{1 - X^2 - Y^2 - Z^2} + 2 \sqrt{6} (1 + 2 h) X^2 \zeta \sqrt{1 - X^2 - Y^2 - Z^2} \right) \nonumber \\
&\quad + Z \left( -3 + 3 (2 + h) X^2 - 3 (1 + h) X^4 + 3 \sqrt{6} X \zeta \sqrt{1 - X^2 - Y^2 - Z^2} + \sqrt{6} X^3 \zeta \sqrt{1 - X^2 - Y^2 - Z^2} \right)
\Big) \nonumber \\
& + 2 Y^2 \Big( 
3 Z^3 
- 2 \sqrt{6} \zeta \sqrt{1 - X^2 - Y^2 - Z^2} 
+ 2 Z^2 \left( 3 X + \sqrt{6} \zeta \sqrt{1 - X^2 - Y^2 - Z^2} \right) \nonumber \\
&\quad - Z \left( 3 + (-3 + 6 h) X^2 + 2 \sqrt{6} X (\lambda_{0} + \zeta) \sqrt{1 - X^2 - Y^2 - Z^2} \right)
\Big)
\Big) ,
\end{align}
\begin{equation}\label{eq2_case1_infinite}
    \frac{d\zeta}{dT}=\frac{\sqrt{3}X\zeta^2}{\sqrt{2}}.
\end{equation}

\begin{figure}[h!]
    \centering
    \includegraphics[width=0.4\textwidth]{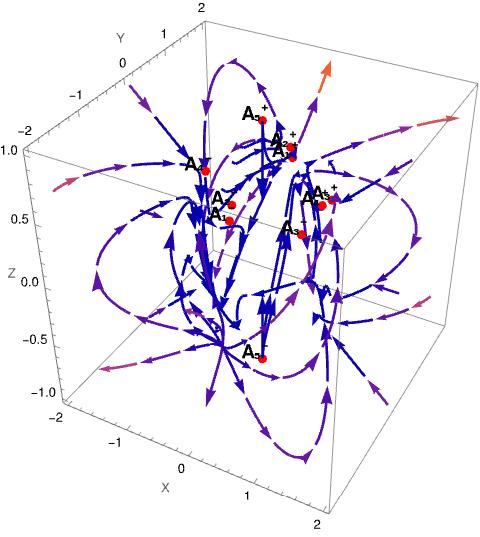}
    \caption{3D phase portrait for the dynamical system given in Eqs. (\ref{eq1_case1_infinite})-(\ref{eq2_case1_infinite}) for $\lambda_0=2,~h=0.5$ (\textbf{Case \ref{case II}}).
   }
    \label{fig4_case2}
\end{figure}

\begin{figure}[h!]
    \centering
    \begin{subfigure}[b]{0.45\textwidth}
        \centering
        \includegraphics[width=\textwidth]{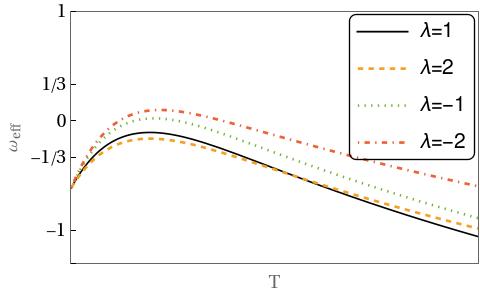}
       \caption{}
        \label{fig:subweff1}
    \end{subfigure}
    \begin{subfigure}[b]{0.45\textwidth}
        \centering
        \includegraphics[width=\textwidth]{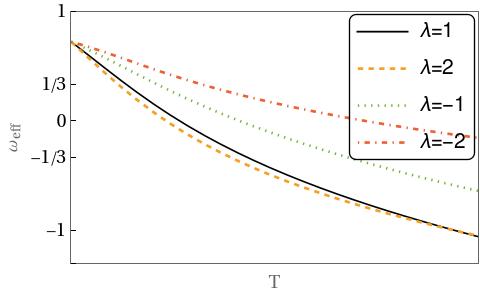}
        \caption{}
        \label{fig:subweff2}
    \end{subfigure}
    
    \caption{Qualitative evolution of the equation of state parameter of the dynamical system (\ref{eq1_case1_infinite})-(\ref{eq2_case1_infinite})
for different values of $\lambda$, with initial conditions ($X[0]=0.5,~Y[0]=0.4 ,~Z[0]=0.3,~\zeta[0]=0.7$). The left plot corresponds to $h=-1$ and the right to $h=1$.
(\textbf{Case \ref{case II}}).}
    \label{fig1:combined}
\end{figure}

\begin{table}[h]
    \centering
    \resizebox{\textwidth}{!}{ 
    \begin{tabular}{|c|c|c|c|c|c|}
        \hline
        Critical point & $(X,Y,Z,\zeta)$ & Existence &$w_{eff}$ & $q$ & Can be Attractor? \\ \hline
        $A_{1}^{\pm}$ & $\left(0,\pm \sqrt{1-Z^2},Z,\zeta\right)$ & $-1 \leq Z\leq 1$ & $\pm \infty$ &  $\pm \infty$ &No\\ \hline
       $A_{2}^{\pm}$ &$\left(0,\pm \sqrt{1-Z^2},Z,0 \right)$ &$-1 \leq Z\leq 1$  & $\pm \infty$ &  $\pm \infty$&No\\ \hline
        $A_{3}^{\pm}$ & $\left( \frac{Z}{h},\pm \frac{\sqrt{h^2-Z^2-h^2Z^2}}{h},Z,0 \right)$ & $h \neq 0 \quad \text{and} \quad -\sqrt{\frac{h^2}{1 + h^2}} \leq Z \leq \sqrt{\frac{h^2}{1 + h^2}}$ & $\pm \infty$ &  $\pm \infty$ &No\\ \hline
          $A_{4}^{\pm}$ & $\left(\pm \sqrt{1-Z^2},0,Z,0\right)$&$-1 \leq Z\leq 1$ & $\pm \infty$ &  $\pm \infty$&No\\ \hline
           $A_{5}^{\pm}$ & $\left(0,0,\pm1,\zeta\right)$&Always &  $0$ & $\frac{1}{2}$&No \\ \hline
         
    \end{tabular}
    }
    \caption{Critical points and their physical properties}
    \label{table2_case2}
\end{table}
From the analysis it is revealed that the stationary points $A_{1}^{\pm}$,  $A_{2}^{\pm}$,  $A_{3}^{\pm}$ and $A_{4}^{\pm}$ describe a cosmological solution with $w_{eff}=\pm \infty$ and $q=\pm\infty$, that is Big Crunch or Big Rip. However, stationary points $A_{5}^{\pm}$ describe the dust fluid with $w_{eff}=0$ and $q=\frac{1}{2}$. Although the detailed stability analysis is not presented here, we conclude that the stationary points at infinity, when they exist, are invariably saddle points. A qualitative analysis has been conducted, supported by the 3D phase portrait shown in Fig \ref{fig4_case2}, which clearly illustrates the unstable nature of the critical points.  Moreover, we present the qualitative evolution of the equation of state parameter (\ref{weffinfinity}), in Fig \ref{fig:subweff1} and in Fig \ref{fig:subweff2}.
    
    
\newpage
\subsection{$\zeta=\zeta_0$ and $\lambda$ is variable}\label{case III}
For this scenario, from (\ref{connIIvar}), we obtain $f=\frac{\zeta_0^2 \phi^2}{4}$. We choose an exponential law for the potential function $U(\phi)=U_0e^{\beta\phi}$, as the power law case has already been analyzed in Subsection \ref{case I}. 
\subsubsection{Finite fixed point analysis}
The autonomous dynamical system is given by
\begin{equation}\label{eq9}
\overline{x}=-\frac{3x}{2}-\frac{3x^2 z}{2}+\frac{3hx^3}{2}-\frac{3xy}{4},
\end{equation}
\begin{equation}
\overline{y}=\sqrt{6}\lambda xy+3y-3yzx+3hx^2 y-\frac{3y^2}{2}+\sqrt{6}xy\zeta_0,
\end{equation}
\begin{equation}
    \overline{z}=\sqrt{6}\zeta_0-\frac{3z}{2}-\sqrt{6}x^2h\zeta_0+\frac{\sqrt{3}\lambda y}{\sqrt{2}}+\sqrt{6}xz\zeta_0-\frac{3z^2 x}{2}+\frac{3hx^2z}{2}-\frac{3yz}{4},
\end{equation}
\begin{align}\label{eq10}
 \overline{\lambda}=\frac{\sqrt{3}x \lambda \zeta_0}{\sqrt{2}}.
\end{align}

\begin{table}[h]
    \centering
    \resizebox{\textwidth}{!}{ 
    \begin{tabular}{|c|c|c|c|c|c|}
        \hline
        Critical point & $(x,y,z,\lambda)$  & Existence &$w_{eff}$ & $q$ & Can be Attractor? \\ \hline
        $P_1$ & $\left( 0,0,2\sqrt{\frac{2}{3}} \zeta_{0} ,\lambda\right)$ & $\lambda=$ arbitrary & $0$ & $\frac{1}{2}$&No \\ \hline
       $P_2$ & $\left( 0,2,z,\frac{1}{2} (\sqrt{6}z-2\zeta_0) \right)$ & $z=$ arbitrary  & $-1$ & $-1$&Yes  \\ \hline
        $P_3$ & $\left( x,0,\frac{-1+hx^2}{x},0 \right)$ & $x \neq 0$ & $1+2\sqrt{\frac{2}{3}}x\zeta_0$ & $2+\sqrt{6}x\zeta_0$ &No \\ \hline
          $P_4$ & $\left( x,0,\frac{-1+hx^2}{x},\lambda \right)$& $\zeta_0 = 0, x \neq 0$,$\lambda=$ arbitrary &$1$& $2$ &No  \\ \hline
            $P_5$  & $\left( x,0,\frac{-1+hx^2}{x},\sqrt{6}(z-hx) \right)$ & $\zeta_0 = 0, x \neq 0$ &$1$& $2$ &No  \\ \hline
             $P_6$  & $\left( -\frac{\sqrt{6}}{\zeta_0},0,\frac{6hx+\sqrt{6}\zeta_0}{6},0 \right)$ & $\zeta_0 \ne 0$  & $-3+\frac{6h}{\zeta_0^2}+\frac{\sqrt{6}hx}{\zeta_0}$ & $-4+\frac{9h}{\zeta_0^2}+\frac{3\sqrt{\frac{3}{2}}hx}{\zeta_0}$&No  \\ \hline
    \end{tabular}
    }
    \caption{Critical points and their physical properties}
    \label{table1_case3}
\end{table}
The detailed examination of the stationary points for the dynamical system (\ref{eq9})-(\ref{eq10})  is summarized in Table \ref{table1_case3},  which specifies the conditions under which they exist, their attractor features, and their related cosmological parameters. 

The critical point $P_1$ represents a dust-filled cosmological phase, as $w_{eff}=0$, $ q=\frac{1}{2}$. The eigenvalues corresponding to this point are $e_{1}(P_1)=3, e_{2}(P_1)=-\frac{3}{2}$, $e_{3}(P_1)=-\frac{3}{2}$, and $e_{4}(P_1)=0$, providing its unstable asymptotic solution. 

The fixed point $P_2$ describes a de Sitter universe as $w_{eff}=-1$ and $ q=-1$. It has stable asymptotic nature as the eigenvalues at this are $e_{1}(P_2)=-3, e_{2}(P_2)=-3$, $e_{3}(P_2)=-3$, and $e_{4}(P_2)=-3$. 

The nature of point $P_3$ is influenced by the parameter $x$ and constant $\zeta_0$, in the limiting scenario where either of them vanishes, the critical point $P_3$ reduces to a configuration corresponding to a stiff fluid, the eigenvalues are $e_{1}(P_3)=-6, e_{2}(P_3)=\sqrt{\frac{3}{2}}x\zeta_0$, $e_{3}(P_3)=6+\sqrt{6}x\zeta_0$, and $e_{4}(P_3)=3+\sqrt{6}x\zeta_0$ and it is unstable for $(x>0 \land \zeta_0 \ge -\frac{\sqrt{6}}{x})$. 

The fixed points $P_4$ and $P_5$ represent the stiff fluid decelerated universe since $w_{eff}=1$ and $ q=2$. Both have an unstable nature depicted by their corresponding eigenvalues that are $e_{1}(P_4)=0, e_{2}(P_4)=0$, $e_{3}(P_4)=3$ and $e_{4}(P_4)=6+\sqrt{6}x\lambda$ also $e_{1}(P_5)=0, e_{2}(P_5)=0$, $e_{3}(P_5)=3$ and $e_{4}(P_5)=-6(-1+hx^2-xz)$. 

The physical characteristics of CP $P_6$ are also parametric dependent and its eigenvalues are  $e_{1}(P_6)=-3, e_{2}(P_6)=\frac{3h(6+\sqrt{6}x\zeta_0)}{\zeta_0^2}$, $e_{3}(P_6)=\frac{3(6h+\sqrt{6}hx\zeta_0)}{2\zeta_0^2}$ and $e_{4}(P_6)=\frac{3(18h+3\sqrt{6}hx\zeta_0-2\zeta_0^2)}{2\zeta_0^2}$ and it is unstable for $(x>0 \land \zeta_0 >0 \land h \ge0)$. 

\subsubsection{Analysis at infinity}
To examine the analysis at infinity for the dynamical system (\ref{eq9})-(\ref{eq10}), we define the Poincare variables
\begin{equation*}
    x=\frac{X}{\rho}, ~~y=\frac{Y}{\rho}, ~~z=\frac{Z}{\rho}.
\end{equation*}
where $\rho= \sqrt{1-X^2-Y^2-Z^2}$ and $dt=\sqrt{1-X^2-Y^2-Z^2} dT$ is the new independent variable. At infinity, the dynamical system can be written as 
\begin{equation}
\begin{aligned}\label{case2_infinite}
\frac{dX}{dT} =& -\frac{1}{4} X \Bigg(
6(1 + h) X^4 - 12 Y^4 - 6 X^3 Z 
+ Y^2 \left(6 - 6 Z^2 - 4 \sqrt{6} \zeta_{0}  Z \sqrt{1 - X^2 - Y^2 - Z^2} \right) \\
& + 2 (-1 + Z^2) \left(-3 + 3 Z^2 - 2 \sqrt{6} \zeta_{0}  Z \sqrt{1 - X^2 - Y^2 - Z^2} \right)- 2 Y^3 \left(3 \sqrt{1 - X^2 - Y^2 - Z^2} + \sqrt{6} Z \lambda \right) \\
& - Y(-1 + Z^2) \left(3 \sqrt{1 - X^2 - Y^2 - Z^2} + 2 \sqrt{6} Z \lambda \right) \\
& + X \Big( (6 - 12 Y^2) Z - 6 Z^3 
+ 4 \sqrt{6} \zeta_{0} Z^2 \sqrt{1 - X^2 - Y^2 - Z^2} 
+ 4 \sqrt{6} Y^2 \sqrt{1 - X^2 - Y^2 - Z^2} (\zeta_{0}  + \lambda) \Big) \\
& - X^2 \Big( 6(2 + h) + (6 - 12h) Y^2 - 6(2 + h) Z^2 
+ 4 \sqrt{6} (1 + h) \zeta_{0}  Z \sqrt{1 - X^2 - Y^2 - Z^2} \\
& \quad + Y \left(3 \sqrt{1 - X^2 - Y^2 - Z^2} + 2 \sqrt{6} Z \lambda \right) \Big)
\Bigg),
\end{aligned}
\end{equation}

\begin{equation}
\begin{aligned}
\frac{dY}{dT} = \frac{1}{4} Y \Big( 
& -6(1 + h) X^4 + 12 Y^4 + 6 X^3 Z  - 2(-1 + Z^2) \left(6 + 3 Z^2 - 2 \sqrt{6} \, \zeta_{0}  \, Z \sqrt{1 - X^2 - Y^2 - Z^2} \right) \\
& + Y^2 \left(-24 + 6 Z^2 + 4 \sqrt{6} \, \zeta_{0}  \, Z \sqrt{1 - X^2 - Y^2 - Z^2} \right)  + 2 Y^3 \left(3 \sqrt{1 - X^2 - Y^2 - Z^2} + \sqrt{6} \, Z \, \lambda \right) \\
& + Y \left(-6 \sqrt{1 - X^2 - Y^2 - Z^2} + 3 Z^2 \sqrt{1 - X^2 - Y^2 - Z^2} \right. \left. - 2 \sqrt{6} \, Z \lambda + 2 \sqrt{6} \, Z^3 \lambda \right) \\
& + 2 X \left(6(-1 + Y^2) Z + 3 Z^3 
- 2 \sqrt{6} \, \zeta_{0}  \, Z^2 \sqrt{1 - X^2 - Y^2 - Z^2} \right. \left. - 2 \sqrt{6} (-1 + Y^2) \sqrt{1 - X^2 - Y^2 - Z^2} (\zeta_{0}  + \lambda) \right) \\
& + X^2 \left(-6 + 12h + (6 - 12h) Y^2 - 6(2 + h) Z^2 
+ 4 \sqrt{6} (1 + h) \zeta_{0}  Z \sqrt{1 - X^2 - Y^2 - Z^2} \right. \\
& \quad \left. + Y \left(3 \sqrt{1 - X^2 - Y^2 - Z^2} + 2 \sqrt{6} Z \lambda \right) \right)
\Big),
\end{aligned}
\end{equation}

\begin{equation}
\begin{aligned}
\frac{dZ}{dT} = \frac{1}{4} \Big(
& -6 Z^5 
+ Z^3 \big( -6(2 + h) X^2 
- 4 \sqrt{6} \, \zeta_{0}  X \sqrt{1 - X^2 - Y^2 - Z^2} + 3(4 + 2 Y^2 + Y \sqrt{1 - X^2 - Y^2 - Z^2}) \big) \\
& + 2 Z^4 \left(3 X + \sqrt{6} \left(2 \zeta_{0}  \sqrt{1 - X^2 - Y^2 - Z^2} + Y \lambda \right) \right)  - 2 \sqrt{6} \Big( (-1 + Y^2)\left(2 \zeta_{0}  \sqrt{1 - X^2 - Y^2 - Z^2} + Y \lambda \right) \\
& \quad + X^2 \left(2(1 + h)\zeta_{0}  \sqrt{1 - X^2 - Y^2 - Z^2} + Y \lambda \right) \Big)  + 2 Z^2 \Big(3 X^3 + X(-3 + 6 Y^2) 
+ \sqrt{6} (-2 + Y^2)\\
& \quad \times \left(2 \zeta_{0}  \sqrt{1 - X^2 - Y^2 - Z^2} + Y \lambda \right) + \sqrt{6} X^2 \left(2(1 + h)\zeta_{0}  \sqrt{1 - X^2 - Y^2 - Z^2} + Y \lambda \right) \Big) \\
& + Z \Big( -6 - 6(1 + h)X^4 - 6 Y^2 + 12 Y^4 
- 3 Y \sqrt{1 - X^2 - Y^2 - Z^2}  + 6 Y^3 \sqrt{1 - X^2 - Y^2 - Z^2}  \\
&
+ 3 X^2 \left(2(2 + h) + (2 - 4h) Y^2 + Y \sqrt{1 - X^2 - Y^2 - Z^2} \right) - 4 \sqrt{6} X \sqrt{1 - X^2 - Y^2 - Z^2} 
(-\zeta_{0}  + Y^2(\zeta_{0}  + \lambda)) \Big)
\Big),
\end{aligned}
\end{equation}
\begin{align}\label{eq2_case2_infinite}
    \frac{d\lambda}{dT}=\frac{\sqrt{3}X \lambda \zeta_0}{\sqrt{2}}.
\end{align}

\begin{figure}[h!]
    \centering
    \includegraphics[width=0.4\textwidth]{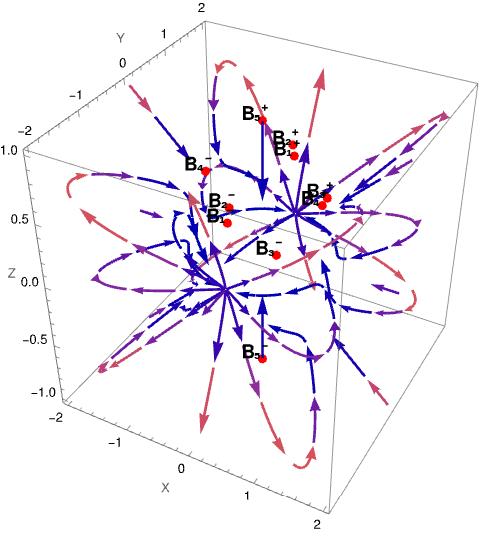}
    \caption{3D phase portrait for the dynamical system given in Eqs. (\ref{case2_infinite})-(\ref{eq2_case2_infinite}) for $\zeta_0=2,~h=0.5$ (\textbf{Case \ref{case III}}).
   }
    \label{fig3_case3}
\end{figure}

\begin{figure}[h!]
    \centering
    \begin{subfigure}[b]{0.45\textwidth}
        \centering
        \includegraphics[width=\textwidth]{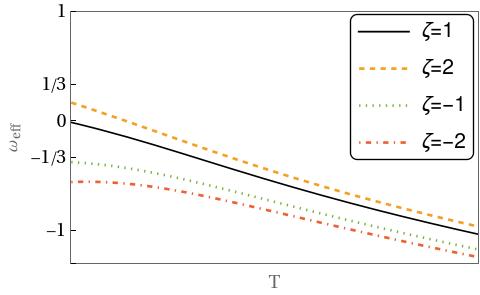}
       \caption{}
        \label{fig:subweff11}
    \end{subfigure}
    \begin{subfigure}[b]{0.45\textwidth}
        \centering
        \includegraphics[width=\textwidth]{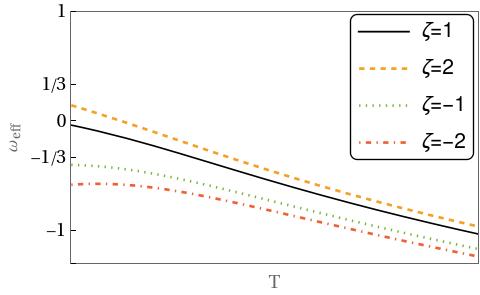}
        \caption{}
        \label{fig:subweff22}
    \end{subfigure}
    
    \caption{Qualitative evolution of the equation of state parameter of the dynamical system (\ref{case2_infinite})-(\ref{eq2_case2_infinite})
for different values of $\zeta$, with initial conditions ($X[0]=0.1,~Y[0]=0.3 ,~Z[0]=0.3,~\lambda[0]=0.7$). The left plot corresponds to $h=1$ and the right to $h=-1$.
(\textbf{Case \ref{case III}}).}
    \label{fig2:combined}
\end{figure}

\begin{table}[h]
    \centering
    \resizebox{\textwidth}{!}{ 
    \begin{tabular}{|c|c|c|c|c|c|c|}
        \hline
        Critical point & $(X,Y,Z,\lambda)$  & Existence &$w_{eff}$ & $q$ & Can be Attractor?\\ \hline
        $B_{1}^{\pm}$ &  $\left(0,\pm \sqrt{1-Z^2},Z,\lambda\right)$&$-1\leq Z\leq 1 $ & $\pm \infty$ &  $\pm \infty$&No\\ \hline
       $B_{2}^{\pm}$ & $\left(0,\pm \sqrt{1-Z^2},Z,0 \right)$&$-1\leq Z\leq 1 $ & $\pm \infty$ &  $\pm \infty$ &No \\ \hline
        $B_{3}^{\pm}$ &  $\left( \frac{Z}{h},\pm \frac{\sqrt{h^2-Z^2-h^2Z^2}}{h},Z,0 \right)$&$h \neq 0 \quad \text{and} \quad -\sqrt{\frac{h^2}{1 + h^2}} \leq Z \leq \sqrt{\frac{h^2}{1 + h^2}} $ & $\pm \infty$ &  $\pm \infty$ &No \\ \hline
          $B_{4}^{\pm}$ &$\left(\pm \sqrt{1-Z^2},0,Z,0\right)$&$-1\leq Z\leq 1 $& $\pm \infty$ &  $\pm \infty$ &No\\ \hline
          $B_{5}^{\pm}$ &$\left(0,0,\pm1,\lambda\right)$&Always &  $0$ & $\frac{1}{2}$&No\\ \hline
    \end{tabular}
    }
    \caption{Critical points and their physical properties}
    \label{table2_case3}
\end{table}
The dynamical analysis demonstrates that the stationary points $B_{1}^{\pm}$,  $B_{2}^{\pm}$, $B_{3}^{\pm}$, $B_{4}^{\pm}$ correspond to cosmological scenarios characterized by $w_{eff}=\pm \infty$ and $q=\pm \infty$,  indicative of extreme future or past singularities such as a Big Rip or Big Crunch. In contrast, the points $B_{5}^{\pm}$ represent a decelerated universe dominated by dust-like matter, with $w_{eff}=0$ and deceleration parameter $q=\frac{1}{2}$. it is evident that the stationary points at infinity, when they exist, consistently behave as saddle points. This conclusion is further supported by a qualitative examination, including the 3D phase space portrait in Fig. \ref{fig3_case3}, which clearly demonstrates the intrinsic instability of these critical points. In addition, the qualitative evolution of the equation of state parameter (\ref{weffinfinity}) is illustrated in Fig \ref{fig:subweff11} and in Fig \ref{fig:subweff22}.

    

\newpage
\subsection{Discussion when $\lambda$ and $\zeta$ both are variable}\label{case IV}
Since natural choices for the coupling and potential functions are the power-law and exponential forms, we have identified four possible scenarios to explore, as previously outlined. Three of these cases have been thoroughly studied in Subsections \ref{case I}, \ref{case II}, and \ref{case III}. The remaining scenario corresponds to the case where both $\lambda$ and $\zeta$  are treated as variables. In this scenario, it follows that  $\zeta=\zeta(\lambda)$ or $\lambda=\lambda(\zeta)$.  


We observe that the stationary points form a continuous family of solutions, exhibiting behavior qualitatively similar to the cases already examined, and the functional form of the potential only constrains the values of the parameter $\lambda$ at the stationary points. As such, the dynamical features do not introduce any fundamentally new physics beyond what has already been captured in our previous analysis. For this reason, we have chosen to omit this scenario from further consideration in this work. 

It is important to mention that the stability properties of the critical points are also affected by the arbitrary functional forms of the theory. 

\section{A brief comparison with scalar-torsion and scalar-tensor frameworks}\label{sec8}
In this section, we provide a brief comparison between our findings and some existing scalar-tensor extension of GR and metric teleparallel gravity. As mentioned above, the non-metricity version of the scalar-tensor theory formulated from the connection class I in FLRW matches exactly with the scalar-torsion theory. So it is interesting to explore the benefit of the extra degree of freedom of connection class II in our study. For instance, in \cite{saridakisST}, the authors explored the cosmological dynamics of teleparallel gravity with a non-minimal coupling between the scalar field and torsion. They adopted a power-law forms for both coupling function and the potential, and also focussed primarily on the case of a quadratic coupling function, closely related to our case \ref{case I}. In their study, a single critical point associated with a de Sitter solution, denoted as $Q_1$, emerged under the condition of a negative coupling constant. However, this point was shown to be unstable. In contrast, our analysis yields a stable de Sitter fixed point, labeled as B, without requiring any specific constraint. Furthermore, while the remaining fixed points in their model exhibited instability and their physical viability was strongly dependent on the coupling constant, our framework offers a broader range of physically meaningful solutions.

The phase-space analysis of a multi-scalar torsion theory was investigated in \cite{comp2}, where the authors considered exponential forms for both coupling function and the potential, an approach analogous to our case \ref{case II}. Their dynamical system yielded five critical points, among which two, denoted as  $P_{2},~P_{3}$, correspond to matter-dominated cosmological solutions, contingent upon specific choices of the model parameters. In comparison, our model admits four critical points, with only one, labeled as A, consistently representing a matter-dominated universe. However, the remaining critical points in their setup were capable of describing accelerated expansion and other cosmologically relevant behaviors, depending on the parameter space. A key distinction between the two analyses lies in the asymptotic behavior: while all critical points approach a Big Rip singularity in the infinite phase-space regime, our model permits two fixed points associated with matter dominance that avoid such singular behavior, offering a potentially more realistic cosmological evolution.

A dynamical analysis involving a canonical scalar field within the framework of teleparallel dark energy was conducted in \cite{comp3}, where an exponential potential was assumed, similar to the setup considered in case \ref{case III}. Their study identified eight fixed points, whereas our analysis reveals six. Notably, they reported the existence of two stable de Sitter attractors under specific constraints on the coupling constant, while our model admits a single de Sitter attractor without requiring any such constraints. Furthermore, their framework produced two dust-dominated critical points, in contrast to the single dust solution found in our analysis. Both approaches identify two stiff fluid critical points. Importantly, their investigation of the asymptotic regime did not produce any viable late-time attractor solutions, a feature shared with our results.

In the classical scalar-tensor theory, a comprehensive dynamical analysis was carried out for both vacuum and matter-dominated scenarios in \cite{comp4}. However, within their framework, the critical point corresponding to the case $n=2$ leads to a non-stable static universe in both regimes. In contrast, our case \ref{case I}  exhibits a much richer dynamical structure, as clearly demonstrated in Table \ref{case1table1}.

In \cite{comp5}, the dynamical system analysis of a non-minimally coupled scalar field was discussed. According to the authors, they could not find any critical point for the case analogous to our case \ref{case II}, see the Table \ref{table1_case2} and \ref{table2_case2} for our analysis.

In \cite{comp6}, the dynamics of a non-minimally coupled scalar field model with exponential potential corresponding to the setup considered in our case \ref{case III}, was investigated. They identified a single fixed point representing a stable de Sitter solution, under the restrictive condition of a constant scalar field and gravitational constant. Whereas our analysis reveals a broader range of cosmological scenarios, including but not limited to a stable de Sitter solution. Combining we can conclude that the scalar-tensor extension of non-metricity gravity, precisely with the connection class II with the extra degree of freedom gives much richer dynamics than the earlier attempts.

\section{Concluding remarks}\label{sec7}
Our intention is to study the recently proposed scalar-tensor extension of the non-metricity gravity theory, specifically its cosmological aspects. Imposing spatial homogeneity and isotropy on a general affine connection with vanishing curvature and torsion yields three classes of connections involving a free time-dependent function $\gamma(t)$. In the first class, however, this function does not affect the cosmological dynamics, reducing the system to the known trivial connection, equivalent to the scalar-torsion case in metric teleparallel gravity. Therefore, in this article, we have considered the second class of connection to investigate the role of $\gamma(t)$, as an extra degree of freedom. We have performed a detailed analysis of the dynamical system starting from the Friedmann field equations. It has been noticed that $\gamma(t)$ always appears in the system multiplied by the derivatives of the coupling function $f(\phi)$, and thus remains totally invisible unless it is a non-minimally coupled theory. We use the Hubble normalization approach to define our dimensionless variables and write the autonomous dynamical system. Based on our variables $\zeta$ and $\lambda$, we have analyzed four possible scenarios. For each case, we have discussed the existence conditions, stability analysis, equation of state parameter $w_{eff}$ and deceleration parameter $q$ of each critical point.

In case \ref{case I}, by considering $\lambda=\lambda_0$ and $\zeta=\zeta_0$, we have obtained a power law form of both coupling and potential function. The dynamical system equations are (\ref{eq1})-(\ref{eq3}). The results for this case are summarized in Table \ref{case1table1}. Fixed point A corresponds to a decelerated, matter-dominated universe and is characterized by instability. Critical point B represents a stable de Sitter solution, indicative of accelerated expansion. In contrast, point C describes a stiff-fluid dominated universe in the limit of vanishing $\zeta_0$, but it remains dynamically unstable.

In case \ref{case II}, we have considered $\lambda=\lambda_0$ and $\zeta$ as variable, and dynamical analysis of the finite fixed points is presented in the system of equations (\ref{4})-(\ref{7}). The detailed stationary point analysis is listed in Table \ref{table1_case2}. Point $A$ describes an unstable, decelerated matter-dominated phase, point $B$ corresponds to a stable de Sitter state, point $C$ yields an unstable stiff fluid universe, and also $D$ leads to an unstable stiff fluid universe in the zero limit of $h$. The asymptotic solution at infinity for this case reveals that the stationary points $A_{1}^{\pm}$,  $A_{2}^{\pm}$,  $A_{3}^{\pm}$ and $A_{4}^{\pm}$ describe Big Crunch or Big Rip. However, stationary points $A_{5}^{\pm}$ describe the dust fluid universe with unstable behavior. The qualitative evolution of $w_{eff}$ (\ref{weffinfinity}) is provided in Fig \ref{fig:subweff1} and in Fig \ref{fig:subweff2}.

In case \ref{case III}, we have considered $\zeta=\zeta_0$ and $\lambda$ as variable, for this case, we have a power law form of the coupling function, and the potential is in exponential form. The dynamical analysis of the finite fixed points is based on the system of equations written in (\ref{eq9})-(\ref{eq10}). The CP $P_1$ gives an unstable, decelerated dust-dominated phase, point $P_2$ describes a stable de Sitter universe, $P_4$ and $P_5$ yield an unstable stiff fluid universe. $P_3$ also gives an unstable stiff fluid universe for $\zeta_0=0$, whereas $P_6$ provides an unphysical scenario. The asymptotic solution at infinity highlights that the critical points $B_{1}^{\pm}$,  $B_{2}^{\pm}$, $B_{3}^{\pm}$, and $B_{4}^{\pm}$ lead towards Big Crunch or Big Rip. But fixed points $B_{5}^{\pm}$ represent the dust fluid universe with unstable behavior.  In addition, the qualitative evolution of the equation of state parameter is given in Fig \ref{fig:subweff11} and in Fig \ref{fig:subweff22}.

In case \ref{case IV}, when both $\zeta$ and $\lambda$ are variables, the dynamical features do not offer new physics beyond our earlier analysis, so we exclude this scenario from further discussion.

In conclusion, although the alternative FLRW connection with an additional degree of freedom might initially appear to introduce potentially problematic effects on the scalar field dynamics, our analysis indicates that it behaves consistently within the studied framework. Nonetheless, further investigations, particularly those incorporating observational data, are necessary to assess whether the influence of this extra degree of freedom can provide useful insights into certain phenomena or poses significant challenges.



\end{document}